\documentclass[twocolumn,fontsize=10pt,DIV=17]{scrreprt}

\usepackage[normalem]{ulem}
\usepackage[mathscr]{eucal}
\usepackage[version=3]{mhchem}

\usepackage{epsfig,latexsym,amsfonts,amsmath,amsthm,amssymb,amsbsy,multirow,slashed,wasysym,textcomp,dsfont,comment,mathtools,cancel,datetime,appendix,BOONDOX-calo}
\usepackage{tocloft}
\usepackage{bbold}
\usepackage{tikz}
\usetikzlibrary{backgrounds}
\usetikzlibrary{patterns}
\usetikzlibrary{arrows.meta}
\tikzstyle{bag} = [align=center]
\usetikzlibrary{decorations.pathmorphing}
\usetikzlibrary{bending}
\usepackage{tikz-feynman}

\tikzfeynmanset{
  fermion1/.style={
    /tikz/postaction={
      /tikz/decoration={
        markings,
        mark=at position 0.5 with {
          \node[
            transform shape,
            xshift=-0.5mm,
            fill,
            dart tail angle=100,
            inner sep=1.3pt,
            draw=none,
            dart
          ] { };
        },
      },
      /tikz/decorate=true,
    },
  },
  fermion2/.style={
    /tikz/postaction={
      /tikz/decoration={
        markings,
        mark=at position 0.5 with {
          \arrow{>[length=6pt, width=5pt]};
        },
      },
      /tikz/decorate=true,
    },
  },
}

\DeclareOldFontCommand{\bf}{\normalfont\bfseries}{\mathbf}

\def\bea{\begin{eqnarray}}
\def\eea{\end{eqnarray}}

\usepackage{hyperref}
\usepackage{subcaption}

 \newcommand{\badat}{\begin{alignedat}}
 \newcommand{\eadat}{\end{alignedat}}
 \newcommand\scalemath[2]{\scalebox{#1}{\mbox{\ensuremath{\displaystyle #2}}}}
 \def\be{\begin{equation}}
\def\ee{\end{equation}}

\def\p{\partial}

\usepackage{xcolor}

\newcommand{\pink}[1]{\textcolor{\pink}{#1}}

\definecolor{dblue}{rgb}{0.2,0.50,0.80}

\def\O{\mathcal{O}}

\def\bz{{\bar z}}
\def\bw{{\bar w}}

\def\bz{{\bar z}}
\def\bw{{\bar w}}

\usepackage{setspace}

\usepackage[retainorgcmds]{IEEEtrantools}
\def\bea{\begin{IEEEeqnarray*}}
\def\eea{\end{IEEEeqnarray*}}

\DeclareFontFamily{OT1}{pzc}{}
\DeclareFontShape{OT1}{pzc}{m}{it}{<-> s * [1.10] pzcmi7t}{}
\DeclareMathAlphabet{\mathpzc}{OT1}{pzc}{m}{it}

\usepackage[sort&compress,square,numbers]{natbib} 
\begin{document}
\chapter*{A Chapter on Celestial Holography}
{\bf S. Pasterski}, Perimeter Institute, Waterloo, ON, CA \\


The Celestial Holography program encompasses recent efforts to understand the flat space hologram in terms of a CFT living on the celestial sphere. A key development instigating these efforts came from understanding how soft limits of scattering encode infinite dimensional symmetry enhancements corresponding to the asymptotic symmetry group of the bulk spacetime. Historically, the construction of the bulk-boundary dual pair has followed bottom up approach matching symmetries on both sides. Recently, however, there has been  exciting progress in formulating top down descriptions using insights from twisted holography. This chapter reviews salient aspects of the celestial construction, the status of the dictionary, and active research directions. This is a preprint version of a chapter prepared for the Encyclopedia of Mathematical Physics 2nd edition~\cite{EncyclopediaOM}.

\section*{Introduction}

The holographic principle provides an important toolkit for understanding quantum gravity. Inspired by black hole thermodynamics, it posits that quantum gravity can be encoded in a lower dimensional non gravitational theory living on the boundary of the spacetime. While a precise top down construction from string theory exists for negatively curved spacetimes, in the form of the AdS/CFT correspondence, extending this to astrophysically relevant contexts has been challenging.

Physics at a broad range of energy scales  -- spanning from particle accelerators to gravitational wave experiments -- can be well approximated by spacetimes with vanishing cosmological constant. However, one comes across various subtleties in trying to generalize AdS/CFT to a flat limit.  For example, the causal structure of the conformal boundary is quite different, with massless fields entering and exiting the spacetime along null infinity. Intriguingly, this complication is accompanied by a rich set of asymptotic symmetries much larger than the Poincar\'e isometry group.

This chapter provides a survey of a program -- Celestial Holography -- that attempts to understand how to apply the holographic principle to asymptotically flat spacetimes, proposing a dual description of scattering as a CFT living on the celestial sphere~\cite{Pasterski:2021rjz,Strominger:2017zoo,McLoughlin:2022ljp,Raclariu:2021zjz,Pasterski:2021raf}. Historically, this venture followed a bottom up construction matching symmetries on both sides of the proposal.  However, a recent collision of ideas with the twistor and twisted holography communities has offered new insights into the origins of certain celestial symmetries, giving us access to certain top down constructions.

This chapter is organized as follows: We start by reviewing the IR origins of the celestial holography proposal, in particular the equivalence between Ward Identities for asymptotic symmetries and soft theorems~\cite{Strominger:2013jfa,Strominger:2013lka,He:2014laa}. We then discuss how to set up the scattering problem in a boost basis~\cite{Pasterski:2016qvg,Pasterski:2017kqt} where these soft charges become currents living on the codimension-2 celestial sphere. After establishing the map between $S$-matrix elements and CCFT correlation functions, we build up the various entries in the holographic dictionary, culminating in the extraction of a $Lw_{1+\infty}$ from the collinear limits of gravitational scattering~\cite{Strominger:2021mtt}. We then close by examining how these efforts connect to different research programs including twistors, twisted holography, Carrollian CFTs, and the conformal collider literature. We will stick to the case of a 4D bulk throughout, though much of the story generalizes to higher dimensions. An overarching goal will be to demonstrate the value of connecting disparate research disciplines, a theme that started retrospectively with the IR triangle and persists in the active research directions.

\section*{Insights from the Infrared}

The Celestial Holography program traces its origins to the observation that the asymptotic symmetries of asymptotically flat spacetime are infinite dimensional enhancements of the global symmetries of Minkowski space. These symmetries have different incarnations in different presentations of the scattering problem. In this section, we will consider the momentum space amplitudes and asymptotic field configurations, in turn, before showing how the Weinberg soft theorem implies a supertranslation Ward identity. This equivalence is an explicit example of a more general pattern that has led us to predict new observables within gravitational waveforms.

\begin{figure}[ht]
\centering
\vspace{-0.5em}
\begin{tikzpicture}[scale=2.1]
\draw[thick] (0,1)--(1,0);
\draw[<->,thick] (0+.25,.5+.25)-- node[above,sloped]{\color{gray}\footnotesize ~~Fourier transform}  (.5+.25,0+.25) ;
\draw[thick] (1,0)-- node[below]{\color{gray}\footnotesize Ward identity} (-1,0) ;
\node[below] (a) at (-.8,-.2) {Asymptotic Symmetries};
\draw[<->,thick] (0-.25,.5+.25)-- node[above,sloped]{\color{gray}\footnotesize vacuum transition~~} (-.5-.25,0+.25) ;
\draw[thick] (-1,0)  --(0,1) node[above]{Memory Effects};
\node[below] (b) at (1,-.2) {Soft Theorems};
\draw[<->,thick] (-.5-.1,0)--(.5+.1,0) ;
   \draw[->,gray] (-.2,.4) arc
    [
        start angle=-180,
        end angle=120,
        x radius=0.2cm,
        y radius =0.2cm
    ] ;
\end{tikzpicture}
\caption{The Infrared Triangle captures a pattern of relations that has been used to identify missing corners and new iterations.
}
\label{fig:IR_triangle}
\end{figure}
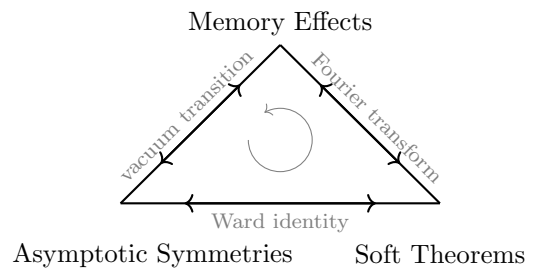

\subsection*{Soft Limits of Scattering}

\begin{figure}[h!]
\centering
\begin{tikzpicture}
   \begin{feynman}[every blob={/tikz/fill=gray!30,/tikz/inner sep=2pt}]
\vertex at (-1,-2)  (a) ;
\vertex at (0,-2)  (b) ;
\vertex at (1,-2)  (c) ;
\vertex at (-1,2)  (d) ;
\vertex at (0,2)  (e) ;
\vertex at (1,2)  (f) ;
 \node at (-.5,-2) {\ldots};
  \node at (.5,2) {\ldots};
  \node at (1,0) {};
\vertex[blob,label={}] (m) at ( 0, 0) {};
\diagram*{
(a) -- [fermion2] (m) -- [boson, edge label={$q^\mu$}] (d);
(b) -- [fermion2] (m) -- [fermion2] (e);
(c) -- [fermion2] (m) -- [fermion2] (f);
};
  \end{feynman}
\end{tikzpicture}
\begin{tikzpicture}
   \begin{feynman}[every blob={/tikz/fill=gray!30,/tikz/inner sep=2pt}]
\vertex at (-1,-2)  (a) ;
\vertex at (0,-2)  (b) ;
\vertex at (1,-2)  (c) ;
\vertex at (-1,2)  (d) ;
\vertex at (0,2)  (e) ;
\vertex at (1,2)  (f) ;
\vertex at (0,1.2)  (g) ;
 \node at (-.5,-2) {\ldots};
  \node at (.5,2) {\ldots};
    \node at (-2,0) {$=$};
       \node at (-2,-.5) {$\omega\rightarrow 0$};
        \node at (2,0) {$+ ~~~$\dots};
\vertex[blob,label={}] (m) at ( 0, 0) {};
\diagram*{
(a) -- [fermion2] (m);
(b) -- [fermion2] (m) -- [fermion2, edge label={$p_i^\mu+q^\mu$}] (g) -- [fermion2] (e);
(c) -- [fermion2] (m) -- [fermion2] (f);
(g) -- [boson] (d);
};
  \end{feynman}
\end{tikzpicture}
\caption{The leading contribution to the amplitude when a graviton or gauge boson goes `soft,' comes from diagrams where this boson couples to each of the `hard' external legs.
\label{soft theorem}}
\end{figure}
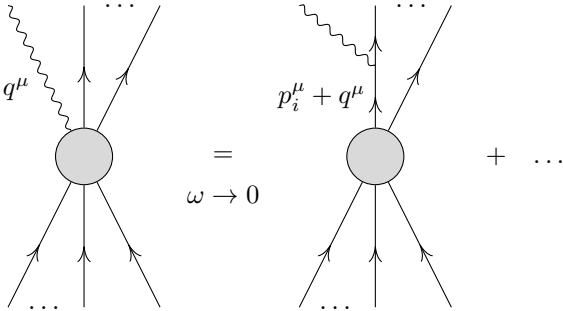

The scattering problem is most often formulated in momentum space, where we can compute $\cal S$-matrix elements perturbatively using Feynman diagrams. Let us consider what happens when the scattering process involves a gauge boson whose energy we take to be very small. In this `soft' limit, the amplitude acquires a pole in the energy. This comes from diagrams where the boson attaches to one of the external legs, as illustrated in figure~\ref{soft theorem}, and the additional propagator is proportional to 
\be
\frac{-i}{(p_i+\eta_i q)^2+m_i^2-i\epsilon}=\frac{-i}{2\eta_i p_i\cdot q-i\epsilon}
\ee
where $\eta_i=\pm1$, depending on whether particle $i$ is outgoing or incoming, and we have used the fact that $p_i$ is on-shell. A seminal result of Weinberg~\cite{Weinberg:1965nx} is that if we take into account numerator factors from the cubic vertex and the spin dependence of the propagator, the soft factor is universal and independent of the spin of the particles to which it couples.

For a soft photon we have the following expansion near $\omega\sim 0$
\be
\langle out|a_\pm(q){\cal S}|in\rangle =e\sum_{i}\eta_i\frac{ Q_ip_i\cdot \epsilon^\pm}{p_i\cdot q}\langle out|\mathcal{S}|in\rangle+...
\ee
where the omitted terms are subleading in $\omega$. Meanwhile for a graviton, the first two orders are universal
\be\label{sfthm}
\langle out|a_{\pm}(q)\mathcal{S}|in\rangle=(S^{(0)\pm}+S^{(1)\pm})\langle out|\mathcal{S}|in\rangle + ... 
\ee
where
\be
S^{(0)\pm}=\frac{\kappa}{2}\sum_i\eta_i\frac{(p_i\cdot \epsilon^\pm)^2}{p_i\cdot q},
\ee
 is the Weinberg leading soft graviton factor, and
\be
S^{(1)\pm}=-i\frac{\kappa}{2}\sum_i\eta_i\frac{p_{i\mu}\epsilon^{\pm\mu\nu}q^\lambda J_{i\lambda\nu}}{p_i\cdot q}
\ee
is the Cachazo-Strominger subleading soft result. Here $\kappa=\sqrt{32\pi G}$. We will now see that this universality has a natural asymptotic symmetry origin.

\subsection*{Asymptotic Symmetries}

\begin{figure}[th!]
\centering
\begin{tikzpicture}[scale=2.3]
\newcommand{\cross}{$\mathbin{\tikz [x=.75ex,y=.75ex,line width=.25ex, blue] \draw (0,0) -- (1,1) (0,1) -- (1,0);}$}
\definecolor{darkgreen}{rgb}{.0, 0.5, .1};
\draw[thick](0,0) --(1,1) node[right] {$i^0$} --(0,2)node[above] {$i^+$}  --(0,0)  node[below] {$i^-$} ;
\draw[red,<->,thick](1+.03,1+.03) -- node[above]{$~u$} (0+.03,2+.03);
\draw[darkgreen,<->,thick](1+.03,1-.03) -- node[below]{$~v$} (0+.03,0-.03);
\draw[thick] (-1+2.5,1.8) circle (.6em);
\draw[blue,fill=blue] (.5,1.5) circle (.75pt);
\draw[blue,fill=blue] (-1+2.5,2.01) circle (.75pt);
\node at (-1+2.5,1.59) {\cross};
\node at (.5,.5) {\cross};
\node[blue] at (-1+2.5, 2.15) {$z$};
\node at (1/2+.3,3/2+.3) {$\cal{I}^+$};
\node at (1/2+.3,1/2-.3) {$\cal{I}^-$};
\end{tikzpicture}
\caption{Penrose diagram for Minkowski space.   Massive particles enter and exit at $i^\pm$ while  massless excitations enter and exit at $\cal{I}^\pm$. There is an $S^2$ over each point besides $r=0$.
\label{penrose}}
\end{figure}
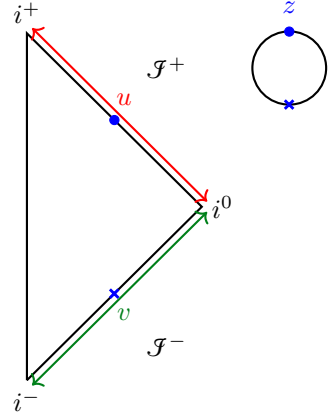

Contemporaneously with Weinberg's soft theorem result, Bondi, Van der Burg, Metzner, and Sachs (BMS) were formalizing an asymptotic analysis of asymptotically flat spacetimes~\cite{Bondi:1962px,Sachs:1962wk,Sachs:1962zza}. These spacetimes are solutions of Einstein's equations with vanishing cosmological constant but non-trivial matter content. Such spacetimes have the same causal structure as Minkowski space, the Penrose diagram for which is illustrated in figure~\ref{penrose}.

In particular massless excitations enter and exit the spacetime along null components of the conformal boundary $\mathcal{I}^\pm$. The aim of the BMS analysis was to identify the free data at null infinity and the relevant asymptotic symmetry group. Upon gauge fixing, the metric near future null infinity can be written as
\be\badat{3}\label{bondi}
ds^2=&-du^2-2dudr+2r^2\gamma_{z\bz}dzd\bz+\frac{2m_B}{r}du^2\\
&+rC_{zz}dz^2+D^zC_{zz}dudz+c.c.\\
&+\frac{1}{r}[\frac{4}{3}N_z-\frac{1}{4}D_z(C_{zz}C^{zz})]dudz+c.c.+...
\eadat\ee
where the omitted terms are subleading at large $r$, holding $u=t-r$ fixed. Solutions to Einstein's equations are then specified by the following free data
\be\label{free}
\{m_B(u_0,z,\bz),N_z(u_0,z,\bz),C_{zz}(u,z,\bz)\}
\ee
where, in particular, the $u$-evolution of the Bondi mass $m_B$ and angular momentum aspect $N_A$ are fixed by the constraint equations $n^\mu [G_{\mu\nu}-8\pi GT_{\mu\nu}]=0$ where $n$ is the null normal to $\mathcal{I}^+$.

The asymptotic symmetry group is identified by looking at residual diffeomorphisms that preserve Bondi gauge~\eqref{bondi} but act non-trivially on the boundary data~\eqref{free}
\be\label{asg}
\mathrm{Asymptotic~Symmetries}=\frac{\mathrm{Allowed~Symmetries}}{\mathrm{Trivial~Symmetries}} .
\ee
These end up being much larger than the Poincar\'e isometries of Minkowski space, instead including angle dependent supertranslations and superrotations
\be\badat{3}\label{xiyf}
\xi&=f\p_u-\frac{1}{r}(D^zf \p_z+D^\bz f \p_\bz)+D^zD_z f\p_r
\\&~~+(1+\frac{u}{2r})Y^z\p_z-\frac{u}{2r}D^\bz D_z Y^z\p_\bz
\\
&~~-\frac{1}{2}(u+r)D_zY^z\p_r +\frac{u}{2}D_zY^z\p_u+c.c.
\eadat
\ee
where $f=f(z,\bz)$ and $Y=Y(z)$.
We will now show that these asymptotic symmetries are indeed symmetries of scattering.

\subsection*{Ward Identities}

In the previous section we examined the asymptotic structure near future null infinity. A similar analysis holds near past null infinity. A key insight of Strominger was to identify a diagonal subgroup of $BMS^+\times BMS^-$ that is a symmetry of the ${\cal S}$-matrix. This subgroup involves an antipodal matching of the symmetry generators across $i^0$. The proof uses the universal soft theorems studied above.

In this section, we will go through the supertranslation example. One can use the constraint equation\footnote{Here we've grouped both the matter stress tensor and terms quadratic in the gravitational field into the quantity $T_{uu}$.}
\be\label{constraints}
\p_u m_B=\frac{1}{4}[D_z^2N^{zz}+D_\bz^2 N^{\bz\bz}]-T_{uu}
\ee
to recast the canonical charge for supertranslations
\be\label{Qfy}
Q^+_f=\frac{1}{8\pi G}\int_{\mathcal{I}^+_-} 2m_B f
\ee
in terms of radiative data
\be
4\pi GQ^+_f=\int_{\mathcal{I}^+_+} \hspace{-.1em} m_B f-\int_{\mathcal{I}^+} \hspace{-.1em} [\frac{1}{4}D_A D_B N^{AB}-T_{uu}]f.
\ee
For purely massless theories we can make the simplifying assumption that $m_B|_{\mathcal{I}^+_+}$ vanishes, though this is not necessary for the conclusion in what follows. Then the antipodal matching of the metric data 
\be\label{antipodal}
    C_{zz}|_{\mathcal{I}^+_-}=    C_{zz}|_{\mathcal{I}^-_+},~~~m_B|_{\mathcal{I}^+_-}=    m_B|_{\mathcal{I}^-_+}, 
\ee
and the supertranslation parameters $f^\pm(z,\bz)$ gives us a relation between the $u$ integral of the news $N_{AB}=\p_u C_{AB}$, the energy radiated through future null infinity -- namely the shear inclusive ANEC $\int du T_{uu}$ -- and the corresponding quantities at past null infinity
\be\label{eq:combo}
\int du [\frac{1}{4}D_A D_B \p_u C^{AB}-T_{uu}] = \mathrm{ antipodal~at~{\mathcal{I}^-}}.
\ee
Here we've used the fact that $f$ can be an arbitrary function on the sphere to reduce the integral to a single generator, i.e. $f=\delta^{(2)}(z-w)$.

If we insert this relation into an ${\cal S}$-matrix element, then the time ordering will place the $\mathcal{I}^+$ operators near the out state and $\mathcal{I}^-$ operators near the in state, giving a relation of the form
\be\label{wardid}
\langle out | Q^+_f \mathcal{S}-\mathcal{S}Q^-_f|in\rangle=0
\ee
where the charges further split into a soft and hard part
\be\label{qsplit}
Q^\pm_f=Q_{S,f}^\pm+Q_{H,f}^\pm.
\ee
The soft part is just the $u$ integral of the news. We will now see why this name is apt. If we look at the perturbative mode expansion of a radiative graviton, in the large-$r$ fixed-$u$ limit, the phase space integral localizes on the celestial sphere
\be
C_{zz}=\frac{-i\kappa}{4\pi^2 (1+z\bz)^2}\int_0^\infty d\omega [a_+e^{-i\omega u}+ a_-^\dagger e^{-i\omega u} ].
\ee
Now we see that the $\int du \p_u C_{AB}$ appearing in~\eqref{eq:combo} precisely picks out the pole in $\omega$ for an additional soft outgoing graviton. Namely, the evaluation of~\eqref{wardid} is the leading soft theorem in~\eqref{sfthm}!

\vspace{1em}

\noindent The above example was only one iteration of a pattern of connections illustrated in figure~\ref{fig:IR_triangle}. By identifying the universal pattern, a brand new iteration was discovered for the subleading soft graviton, including a new experimental prediction: the spin memory effect~\cite{Pasterski:2015tva}. The asymptotic symmetry that spurred these new additions was the superrotations in~\eqref{xiyf}, which enhance the Lorenz group to two copies of the Witt algebra. This proposed Virasoro symmetry of scattering gave a tantalizing hint at a codimension-2 description of the ${\cal S}$-matrix in terms of conformal correlation functions on the celestial sphere, to which we now turn.

\section*{Celestial Amplitudes}

The analog of the soft charge in~\eqref{qsplit} for the superrotation symmetry can be manipulated into a candidate stress tensor for a 2D CFT living on the celestial sphere~\cite{Cheung:2016iub,Kapec:2016jld}. Here, the weights of operators are given by Rindler energies, which are not diagonalized in the standard momentum space basis. This suggests that if we would like to take better advantage of the infinite dimensional symmetry enhancements encoded in the soft limits of scattering, we should recast our ${\cal S}$-matrix elements in a basis that diagonalizes these collinear boosts. We call these objects celestial amplitudes.

\subsection*{Conformal Primary Wavefunctions}

While the asymptotic symmetry enhancements motivate the celestial hologram, if we stick to the global symmetry group the statement becomes quite simple: boosts in 4D act as global conformal transformations on the 2D celestial sphere. Accordingly, we can recast our scattering amplitudes into a basis that transforms under the Lorentz group like 2D quasi-primaries
\begin{equation}
\langle out|S|in\rangle_{boost}=\langle\mathcal{O}_{\Delta_1,J_1}(z_1,\bz_1)...\mathcal{O}_{\Delta_n,J_n}(z_n,\bz_n) \rangle 
\label{eq:4Ddict}
\end{equation}
by preparing an appropriate set of wavepackets for the external scattering states~\cite{deBoer:2003vf,Pasterski:2016qvg,Pasterski:2017kqt}.

\vspace{1em}
 
\noindent{\bf Definition:} A \textit{conformal primary wavefunction} is a function on $\mathbb{R}^{1,3}\times\mathbb{C}$ which transforms under $SL(2,\mathbb{C})$  as a 2D conformal primary of dimension $\Delta$ and spin $J$, and a 4D field of spin-$s$.
\begin{equation}\label{Defgenprim}
\scalemath{0.96}{    \badat{2}
&\Phi^{s}_{\Delta,J}\Big(\Lambda^{\mu}_{~\nu} X^\nu;\frac{a w+b}{cw+d},\frac{{\bar a} \bw+{\bar b}}{{\bar c}\bw+{\bar d}}\Big)\\
&=(cw+d)^{\Delta+J}({\bar c}\bw+{\bar d})^{\Delta-J}D_s(\Lambda)\Phi^{s}_{\Delta,J}(X^\mu;w,\bw).
\eadat}
\end{equation}
We will further demand that these wavefunctions satisfy the appropriate spin-$s$ linearized equation of motion.

\vspace{1em}

 The infinitesimal version of~\eqref{Defgenprim} is just
 \begin{equation}
\label{covariance_generators}
(M_{\mathbb{R}^{1,3}}^{\mu\nu}+M_{\mathbb{C}}^{\mu\nu})\Phi^s_{\Delta,J}(X^\sigma;w,\bw)=0. \, 
\end{equation}
As such, the operators prepared by taking an inner product between these wave packets and a spin-$s$ bulk operator
 \be\label{qdelta}
\O^{s,\pm}_{\Delta,J}(w,\bw)\equiv i(\hat{O}^{s}(X^\mu),\Phi^s_{\Delta^*,-J}(X_\mp^\mu;w,\bw))_{\Sigma}\,
\ee
are designed to transform as 2D quasi-primaries. Here, the $\pm$ superscript indicates whether the operator is incoming or outgoing, which is selected by the analytic continuation $X_\pm^\mu=X^\mu\pm i\{-1,0,0,0\}$. 

If we want to prepare the initial and final states within time ordered correlation functions, it is convenient to push the Cauchy slice to the conformal boundary. This amounts to a flat space version of the extrapolate holographic dictionary. For massless fields, a saddle point analysis similar to the soft theorem discussion for a massless scalar would give
\begin{equation}
     \mathcal{O}_\Delta (z,\bar{z})\equiv \int^{\infty}_{-\infty} d u\, (u-i\epsilon)^{-\Delta}   \lim_{r\to\infty}r\Phi(u,r,z,\bar{z})\,,
     \label{eq:4Ddict0}
\end{equation}
with analogous expressions for spinning fields and where the measure would be replaced by $u-i\epsilon \mapsto v+i\epsilon$ to prepare the incoming states.

We see that this extrapolate dictionary involves two steps. First, the power of $r$ we need to strip out is fixed by the bulk scaling dimension, much like in AdS$_4$/CFT$_3$. Then, the $u$ integral is a `dimensional reduction' to the celestial sphere. Because we are trading one continuous parameter for another, we expect the radiative data for each bulk field to have a continuous spectrum 
\be
\Delta=1+i\lambda~~~~\lambda\in \mathbb{R}
\ee
on the principal series. This establishes the celestial holographic dictionary for the case of a perturbative bulk. While~\eqref{eq:4Ddict0} shows how to go from the position space correlators to the boost basis, it will be more convenient to recast the holographic map in terms of the original momentum space amplitudes, where we have a better understanding of the objects we would like to translate to the celestial basis.

\vspace{1em}

\noindent Before proceeding, we note that we have avoided a discussion of the massive case in this chapter. The conformal primary wavefunction construction still holds, though the position and momentum space integration kernels take a different form.

\subsection*{Amplitudes in a Boost Basis}

Now the celestial primary wavefunctions~\eqref{Defgenprim} take a simple form in terms of the tetrad
\be\label{tetrad}\badat{3}
&l^\mu=\frac{q^\mu}{-q\cdot X}\,, ~~~n^\mu=X^\mu+\frac{X^2}{2}l^\mu\,,\\
&m^\mu=\epsilon^\mu_++(\epsilon_+\cdot X) l^\mu\,, ~~~\bar{m}^\mu=\epsilon^\mu_- +(\epsilon_-\cdot X) l^\mu\,,
\eadat\ee
where
\be
q^\mu=(1+z\bz,z+\bz,i(\bz-z),1-z\bz)
\ee
is a reference null vector while $\sqrt{2}\epsilon_+=\p_{w} q$ and $\sqrt{2}\epsilon_-=\p_{\bar w} q$ are polarization vectors. Namely
\be\begin{array}{ll}\label{mellinCPWs}
     A_{\Delta,J=+ 1}=m \varphi^{\Delta} \,,&~~~ A_{\Delta,J=- 1}=\bar{m} \varphi^{\Delta}\,, \\
    h_{\Delta,J=+2}=m m \varphi^{\Delta} \,,&~~~  h_{\Delta,J=-2}=\bar{m} \bar{m} \varphi^{\Delta}\,,
\end{array}
\ee
where
\be
\varphi^\Delta =\frac{1}{(-q\cdot X)^\Delta}
\ee
is a scalar primary wavefunction.

Up to an overall normalization, this wavefunction is just a Mellin transform of the plane wave
\be\badat{3}
&\frac{1}{(-q\cdot X_\pm)^\Delta}=\frac{(\pm i)^\Delta}{\Gamma(\Delta)}\phi^{\Delta,\pm},\\
&\phi^{\Delta,\pm}=\int_0^\infty d\omega \omega^{\Delta-1}e^{\pm i\omega q\cdot X-\omega\epsilon}
\eadat\ee
and indeed the spinning analogs are Mellin transforms of the corresponding momentum space plane waves, up to additional gauge transformations. 

This gives us a simple definition of the celestial amplitude for massless particles
\be\label{mellin}\badat{3}
\langle \mathcal{O}^\pm_{\Delta_1}(z_1,\bz_1)...\mathcal{O}^\pm_{\Delta_n}(z_n,\bz_n)\rangle~~~~~~~~~~~~~~~~~~\\
=\prod_{i=1}^n \int_0^\infty d\omega_i \omega_i^{\Delta_i-1} \langle out|\mathcal{S}|in\rangle
\eadat\ee
where the 2D spins match the 4D helicities. With this we can begin translating between bulk scattering and celestial correlation functions by performing this transform on features of amplitudes~\cite{Arkani-Hamed:2020gyp}.

\section*{Holographic Dictionary}

We will now proceed to use the map~\eqref{mellin} to fill in entries of the holographic dictionary.

\subsection*{Soft Charges as Currents}

A natural place to start is with the asymptotic symmetry generators that we used to motivate the celestial basis in the first place~\cite{Donnay:2020guq}. Above we claimed that the subleading soft theorem gave a candidate 2D stress tensor. More generally, the Ward identities for the asymptotic symmetries in 4D can be recast as conformal Ward identities for 2D currents by picking an appropriate symmetry parameter.

Consider a tree level amplitude series expanded around the soft limit of one of the external legs\footnote{
Loop level effects will modify the analytic structure in the $\Delta$ plane, introducing higher order poles. However, the BMS symmetries discussed in this section still survive.
}
\be\label{aexp}
A:=\langle out|\mathcal{S}|in\rangle\sim\omega^{-1} A^{(1)}+A^{(0)}+...
\ee
Upon performing a Mellin transform, the coefficients of the powers in $\omega$ turn into residues at integer conformal dimensions on and to the left of the principal series 
\be\label{alim}
\lim\limits_{\Delta\rightarrow-n}(\Delta+n)\int_0^\infty d\omega \omega^{\Delta-1}\sum_k\omega^k A^{(k)}=A^{(n)},
\ee
namely at $\Delta=1,0,-1,-2...$.  While most of these modes are off of the principal series, translations along $u$ induce a shift $\Delta\rightarrow\Delta+1$ in the the conformal dimension of our single particle states, so we are forced to contend with a spectrum analytically continued to complex-$\Delta$. Going back to the soft gravitons studied above, we can construct two holomorphic currents~\cite{Fotopoulos:2019vac}
\be\badat{3}
&P(z)=\lim\limits_{\Delta\rightarrow1}(\Delta-1)\p_{\bz}\mathcal{O}_{\Delta,+2}(z,\bz),\\
&T(z)= \frac{3!}{2\pi}\int d^2w\frac{1}{(z-w)^4} \lim\limits_{\Delta\rightarrow0}\Delta\mathcal{O}_{\Delta,-2}(w,\bw)
\eadat\ee
generating the supertranslation and superrotation symmetries, respectively.

Defining the commutator of of two holomorphic operators as follows
\be\label{rcom}
[A,B](z)=\frac{1}{2\pi i}\oint_z dw A(w)B(z)
\ee
we see that  the mode operators
\be\label{poincarelaurent}
P_{a,-1}=\oint dz z^{a+1}P(z),~~~L_{n}=\oint dz z^{n+1}T(z)
\ee
obey the BMS algebra
\begin{gather}
  \label{BMSalg}
[L_m,L_n]=(m-n)L_{m+n},\\ 
[L_{n},P_{a,b}]=\left(\frac{n-1}{2}-a\right)P_{a+n,b},
\end{gather}
and similarly for the $\bar L_n$. Namely, while our conformal primary wavepackets ensured the scattering amplitudes transformed covariantly under the global conformal symmetries (and Poincar\'e for that matter), insertions of additional collinear graviton states fill out BMS multiplets.

\subsection*{Collinear Limits as OPEs}

If we want to take the 2D hologram seriously, the goal would be to replace Feynman diagrammatic expansions with celestial operator product expansions. Two operators approaching each other on the celestial sphere are going collinear in both position and momentum space. We should thus be able to extract celestial OPEs from the collinear limits of scattering~\cite{Pate:2019lpp}, as we now demonstrate.

We will again restrict ourselves to tree level scattering throguhout. The limit in which two particles go collinear can be captured by the collinear splitting function
\be
\lim\limits_{z_{ij}\rightarrow0}A_{s_k}(p_k)\longrightarrow\sum_{s\in\pm2}\mathrm{Split}^s_{s_is_j}(p_i,p_j)A_{\widehat s_k}(\widehat{p}_k)
\ee
where the amplitude on the right hand side has particles $i$ and $j$ replaced with a spin-$s$ particle with
\be
P^\mu=p_i^\mu+p_j^\mu,~~~\omega=\omega_i+\omega_j.
\ee
The collinear splitting factors are nonzero for the spin configurations $(s,s_i,s_j)=(2,2,2)$, $(-2,2,-2)$, and similarly with all the signs flipped. In particular
\be
\mathrm{Split}^2_{22}(p_i,p_j)=-\frac{\kappa}{2}\frac{\bz_{ij}}{z_{ij}}\frac{\omega^2}{\omega_i\omega_j}.
\ee
Upon performing a change of variables
\be
\omega_i=t\omega,~~~\omega_j=(1-t)\omega,
\ee
and Mellin transforming we get
\be\scalemath{1}{\badat{3}
&-\frac{2}{\kappa}\frac{z_{ij}}{\bz_{ij}}\int_0^\infty d\omega_i\omega_i^{\Delta_i-1}\int_0^\infty d\omega_j\omega_j^{\Delta_j-1}\mathrm{Split}^2_{22}\left(\cdot\right)\\
&=\left[\int_0^1 \hspace{-.2em}  dt t^{\Delta_i-1}(1-t)^{\Delta_j-2}\right]\int_0^\infty \hspace{-.2em} d\omega \omega^{\Delta_i+\Delta_j-1}\left(\cdot\right).
\eadat}\ee
Importantly, since the only $t$-dependence is in the first factor, this integral can be evaluated explicitly. The remaining $\omega$ integral simply takes the lower point amplitude back to the celestial basis for the leg that's been replaced. In all, we get
\be\badat{3}\label{hhope}
&\mathcal{O}_{\Delta_1,+2}(z_1,\bz_1)\mathcal{O}_{\Delta_2,+2}(z_2,\bz_2)\\
&\sim-\frac{\kappa}{2}\frac{\bz_{12}}{z_{12}}B(\Delta_1-1,\Delta_2-1)\mathcal{O}_{\Delta_1+\Delta_2,+2}(z_2,\bz_2).
\eadat\ee
We will next see that from this OPE we can extract an even richer symmetry structure.

\subsection*{Towers of Symmetries}
We've seen above that if we trust the 2D CFT picture we can extract both the BMS symmetries in terms of celestial currents and a celestial operator product from the collinear limits. How far can we push this~\cite{Crawley:2021ivb,Cotler:2023qwh,Freidel:2022skz}? Note that our discussion of soft limits of scattering indicated that there exists a tower of poles on and to the left of the principal series. Similarly the beta function in~\eqref{hhope} ensures that the OPE closes on the residues
\begin{equation}\label{eq:resH}
    H^k(z,\bz) ~:=~ \lim_{\epsilon\to 0}\,\epsilon\,{\cal O}_{k+\epsilon,2}(z,\bz),
\end{equation}
for $k ~=~2, 1,0,-1,...$, where the $k=2$ term is zero for tree level amplitudes. From the global $SL(2,\mathbb{C})$ algebra
\be
\badat{3}
    [ \bar L_{1}, (\bar L_{-1})^k]
     &=k (\bar L_{-1})^{k-1} (2\bar L_0+k-1),
\eadat
\ee
we see that
\begin{equation}
\label{condition:primary_descendants}
    \bar L_{1} (\bar L_{-1})^k |h,\bar h \rangle = k (2\bar h+k-1) (\bar L_{-1})^{k-1} |h,\bar
    h \rangle \, 
\end{equation}
so that the celestial primary states have primary descendants at
\be
\bar h=\frac{1}{2}(\Delta-J)=0,-1,-2, \ldots
\ee
precisely at the residues~\eqref{eq:resH}. Assuming the primary descendants decouple allows us to perform a finite expansion of these currents in the $\bz$ coordinate 
\begin{equation}
    H^k(z,\bz) ~=~  \sum_{m=\frac{k-2}{2}}^{\frac{2-k}{2}}\,\bz^{-\frac{k-2}{2}-m}\,H^k_m(z),
    \label{eq:hkmode}
\end{equation} 
leaving us with the holomorphic currents $H^k_m(z)$. It is worth noting that this truncation is consistent with~\eqref{hhope}, despite the fact that the hermitian conjugation properties of the Lorentz generators do not force primary descendants to be null. Rescaling the generators 
\begin{equation}
    w^p_n ~=~ \frac{1}{\kappa}\,(p-n-1)!(p+n-1)!\,H_n^{-2p+4}~,
\end{equation}
and evaluating the commutator~\eqref{rcom} for each of these modes gives the loop algebra \begin{equation}
    \Big[w^p_n, w^q_m\Big](z) ~=~ \Big[ n(q-1) - m(p-1)\Big]\,w^{p+q-2}_{m+n}(z)~,
    \label{eq:wloop}
\end{equation}
for $p=1,\frac{3}{2},2,\frac{5}{2}...$ and restricted to the wedge $1-p\le m\le p-1$. Namely, the celestial OPE encodes an $ \wedge L  w_{1+\infty}$ symmetry algebra~\cite{Strominger:2021mtt}!

\vspace{1em}

\noindent As for the leading soft theorem examples, this symmetry tower has natural generalizations to gauge fields as well. While there are a plethora of active directions in which one can fill in technical details of the celestial dictionary, we will close by taking a step back to focus on the bigger picture of how various  attempts to understand the flat hologram converge.

\section*{Merging Frameworks}

With the IR triangle, the celestial holography program had its roots in connecting disparate studies from gravitational wave physics, quantum field theory, and asymptotia. Similarly, much of the recent excitement has centered on the variety of interpretations of the celestial symmetry algebras that arise from different research disciplines.

\subsection*{Twistors and Twisted Holography}
The $w_{1+\infty}$ symmetry we encountered in the prior section has a natural twistorial description in terms of symmetries of self dual gravity~\cite{Adamo:2021lrv,Bu:2022iak}. Related twistor space constructions also appear in twisted holography~\cite{Costello:2022wso}. Fleshing out these connections led to an identification of the celestial symmetry algebras within that framework and, subsequently, a first top down construction of a celestial hologram~\cite{Costello:2022jpg}. Thus we have come full circle: from the bottom up origins where we were only matching symmetries, to a full top down construction containing those symmetries.

\subsection*{Carrollian CFTs and Conformal Colliders}

Another interesting set of directions where cross fertilization has occurred emerge from uplifting the celestial construction from 2D back to 3D and 4D. The position space picture of massless scattering in terms of correlation functions of operators at null infinity lends itself to identifying intrinsically defined Carrollian field theory duals living at null infinity~\cite{Donnay:2022aba}. Moreover, the manner in which the celestial operators are smeared along null infinity matches certain light ray operators common in the CFT literature, restricted to a light sheet in 4D~\cite{Hu:2022txx}. In the perturbative limit, these light ray operators further connect to phase space realizations of the celestial symmetry algebras~\cite{Freidel:2021ytz}.  In sum, both the 3D and 4D CFT perspectives give us a better handle on the symmetries of the celestial theory and guide us beyond certain limitations of the standard construction of celestial amplitudes in terms of Mellin transforms of low point momentum space amplitudes.

\subsection*{Further Applications}
Above, we focused on examples of closely connected fields that are interested in the same symmetry generators for different reasons. These connections are especially meaningful because of the fact that different realizations of the same objects give depth to how we understand them. However, this discussion only touches on the interesting applications of the celestial framework.

Our discussion of the IR triangle only briefly mentioned the third corner: memory effects. These low energy observables have, for some iterations, provided new experimental predictions~\cite{Nichols:2017rqr}. For the gravitational case, measuring them would serve as a test of the validity of the approximations of asymptotic flatness built into our framework. While fixing the BMS frame at infinity serves the practical role of ensuring the waveforms one uses obey the requisite conservation laws~\cite{Mitman:2021xkq}, understanding the horizon analog can have interesting theoretical implications for black hole evaporation~\cite{Hawking:2016msc}.

\vspace{1em}
\noindent 
In sum, the main idea at the core of the Celestial Holography program is to understand how to incorporate the holographic nature of quantum gravity -- informed by explorations of black hole physics concretely realized in string theory -- into the frameworks we are using to describe our 3+1 dimensional physical world. Framed in this practical manner, it is hardly surprising that multiple research ventures share this objective. What's truly exciting is that they are actively colliding.

\vspace{-.2em}

\section*{Acknowledgements}

The research of SP is supported by the Celestial Holography Initiative at the Perimeter Institute for Theoretical Physics and by the Simons Collaboration on Celestial Holography. Research at the Perimeter Institute is supported by the Government of Canada through the Department of Innovation, Science and Industry Canada and by the Province of Ontario through the Ministry of Colleges and Universities.

\vspace{-.2em}

\section*{Further Reading}

\begingroup
\def\chapter*#1{}
{\footnotesize
\setstretch{.9}
\bibliographystyle{utphys}
\bibliography{chapter_v2}
}
\endgroup

\end{document}